\documentclass[10pt,letterpaper]{article}
\usepackage[top=0.85in,left=2.75in,footskip=0.75in]{geometry}

\usepackage{amsmath,amssymb}
\usepackage{changepage}
\usepackage{textcomp,marvosym}
\usepackage{cite}
\usepackage{nameref,hyperref}
\usepackage[right]{lineno}
\usepackage[nopatch=eqnum]{microtype}
\DisableLigatures[f]{encoding = *, family = * }
\usepackage[table]{xcolor}
\usepackage{array}

\newcolumntype{+}{!{\vrule width 2pt}}
\newlength\savedwidth

\raggedright
\usepackage[aboveskip=1pt,labelfont=bf,labelsep=period,justification=raggedright,singlelinecheck=off]{caption}

\makeatletter
\renewcommand{\@biblabel}[1]{\quad#1.}
\makeatother

\usepackage{lastpage,fancyhdr,graphicx}
\usepackage{epstopdf}
\fancyheadoffset[L]{2.25in}
\fancyfootoffset[L]{2.25in}
\usepackage[normalem]{ulem}

\usepackage{array}
\usepackage{tabularx}
\usepackage{booktabs}
\usepackage{float}
\begin{document}
\vspace*{0.2in}

\begin{flushleft}
{\Large
\textbf\newline{Process bigraphs and the architecture of compositional systems biology}
}
\newline
\\
Eran Agmon\textsuperscript{1*},
Ryan K.~Spangler\textsuperscript{1}
\\
\bigskip
\textbf{1} Center for Cell Analysis and Modeling, University of Connecticut Health, Farmington, Connecticut, USA
\\
\bigskip
* agmon@uchc.edu
\end{flushleft}

\section*{Abstract}
Building multiscale biological models requires the integration of independently developed submodels, which means moving shared variables between them and coordinating when each submodel runs.
Existing tools typically address isolated biological mechanisms with specific numerical methods, rarely specify which variables each model reads and writes, how those variables are translated, and how model updates are synchronized.
We present Process Bigraph, a framework for composing and simulating multiscale models. 
Process Bigraph generalize the architectural ideas previously used in the Vivarium software into a shared specification that describes process interfaces, hierarchical data structures, composition patterns, and orchestration patterns. 
The paper explains how the framework is organized and why it makes biological models easier to understand, reuse, and build on; the Supplementary Materials provides the formal specification.
We introduce Vivarium 2.0 as an open-source implementation of the process bigraph framework. 
We demonstrate its utility with Spatio-Flux, a standalone library of microbial ecosystem simulations that combine kinetic equations, dynamic flux balance analysis, and different spatial processes.
These simulations illustrated how the framework effectively integrates diverse biological processes, leading to emergent spatially organized population dynamics.
Finally, we discuss implications for emerging multiscale modeling standards.

\textbf{Availability and implementation:} 
Vivarium 2.0 is a suite of software libraries that include:  
(1) bigraph-schema, for defining and operating on data types within a hierarchical JSON-based format; 
(2) process-bigraph, for defining process interfaces, composite simulations, and executing them;  
and (3) viva-superpowers, for AI-assisted construction and composition of process bigraphs.
Spatio-Flux is a reference application used in this paper to demonstrate the framework's utility.
A Supplementary Materials provide detailed descriptions of these software libraries.
All software is open-source and available at: \url{https://github.com/vivarium-collective}.

\section*{Author keywords}
Process Bigraph; Vivarium 2.0; multiscale modeling; model composition; simulation framework


\section*{Introduction}

Systems biology requires infrastructure that enables the free composition of heterogeneous datasets, models, and methods into unified multiscale simulations.
Whereas models in systems biology traditionally focus on the structure or dynamics of a particular subsystem, formulated with a specific method and calibrated under controlled conditions, \textit{compositional systems biology} aims to connect these models, asking critical questions about \textit{the space between models}: 
What variables should a submodel expose through its interface? 
How do coupled models connect and translate across scales? 
How can distinct modeling methods be coupled and translated across scales?
How can models grounded in different biological or physical disciplines communicate to synthesize new knowledge?
Can we build flexible software that integrates diverse datasets and submodels into consistent and reproducible composite simulations?
How can communities of researchers collaboratively access, recombine, and refine models into ever more successful forms?

Systems biology models have not been compositional in the same way as their biological counterparts --- this is in part because models do not have chemistry or physics to mediate their interactions, and require a model for every natural process.
It is true that many modeling paradigms such as ordinary differential equations (ODEs), flux balance analysis (FBA), Bayesian networks, rule-based models, and Markov models are internally compositional in that they can be expanded by adding new expressions or nodes to represent reactions, species, pathways, etc. 
But different aspects of biology require different representations and are typically studied with different modeling paradigms, and these too need to be brought together. 

The need to connect models across scales has driven interest in hybrid approaches that combine modeling paradigms --- such as stochastic with deterministic \cite{haseltine2005origins}, kinetic ODEs with steady state FBA \cite{mahadevan2002dynamic,covert2008integrating}, particle-based with continuous spatial \cite{schaff2016numerical}, COMETS for dynamic FBA with spatial diffusion \cite{dukovski2021metabolic}, and whole-cell models with many interacting processes \cite{karr2012whole,macklin2020simultaneous,maritan2022building,thornburg2022fundamental,stevens2023molecular}.
Recent approaches also integrate mechanistic models with machine-learning \cite{rackauckas2020universal}, enabling data-driven refinement of dynamical systems while retaining mechanistic interpretability.
Multi-cellular models often use agent-based models (ABMs) \cite{walpole2013multiscale,ghaffarizadeh2018physicell,swat2012multi} --- a class of hybrid models that spans two scales, with a model for an environment and models for individual agents. 
However, these are usually tightly coupled with specific software environments or modeling methodologies --- the lack of interoperability across domains limits model reuse, scalability, and collaboration.

To address these challenges, we need a shared interface and runtime protocol—a standard way to define how models expose variables, how data flows between them, and how execution is coordinated.
Such a framework would support mix-and-match models built with different methods, enable reuse and substitution through compatible interfaces, and provide a means to dynamically structure and orchestrate simulations across multiple scales.
Just as communication protocols in distributed systems allow diverse components to interoperate via standardized interfaces and message formats, a composition protocol would allow multiscale biological simulations to be constructed from interoperable parts, validated independently, and integrated into larger models.
This would complement existing standards developed by the COMBINE community—SBML \cite{keating2020sbml}, CellML \cite{lloyd2004cellml}, and SED-ML \cite{waltemath2011reproducible}—and execution registries such as BioSimulators \cite{shaikh2022biosimulators}, by focusing not on unifying models into a single file but on orchestrating heterogeneous models and methods.

With Vivarium~1.0 \cite{agmon2022vivarium}, we implemented these ideas in Python using a framework of processes that exchange data through shared, hierarchically structured stores.
The design supported modularity by allowing processes to be swapped or recombined through explicit interfaces, hierarchy through recursive nesting of processes and stores, and dynamic orchestration coordinating processes operating on different timescales.
This enabled flexible coupling of stochastic and deterministic submodels, synchronization of hybrid processes, and runtime restructuring of simulations.
Vivarium was successfully applied to diverse biological problems, including metabolic and regulatory integration in whole-cell simulations of \textit{E.~coli} \cite{skalnik2023whole}, bacterial chemotaxis integrating motility and intracellular signaling \cite{agmon2020multi}, and tumor morphogenesis combining agent-based models with multiplexed spatial imaging data \cite{hickey2024integrating}.
While these applications demonstrated the framework’s flexibility and scalability, they also revealed that encoding compositional semantics directly in Python limited transparency, interoperability, and reuse.

In particular, model structure, typing, and execution semantics were implicit in code rather than explicitly defined.
This motivated the development of Process Bigraph as a more general and implementation-independent composition protocol.

This paper introduces the Process Bigraph framework, a compositional modeling protocol in which individual \emph{process bigraphs} are formal objects that define how subsystems expose interfaces, connect to one another, and are orchestrated across scales.

Process Bigraph makes compositional semantics explicit by introducing a schema and type system that separate model structure from state and define how updates are validated and applied.
They represent processes, wiring, and hierarchy in a serializable, language-agnostic JSON format, enabling models to be stored, exchanged, and executed across environments.
They also define explicit execution protocols that decouple process specification from runtime context, supporting local, parallel, and distributed execution.
Within this framework, the Composite is formalized as the execution engine, unifying orchestration patterns such as multi-timestepping, workflows, and structural updates.

We present the core conceptual principles together with a graphical language because we believe they provide powerful conceptual tools for reasoning about multiscale biology and what it means to build compositional models as a research community; the Supplementary Materials provides the corresponding formal semantics.
The Process Bigraph framework is implemented in Vivarium~2.0, an open-source successor to Vivarium~1.0 that adopts a language-agnostic exchange format in JSON.
The Vivarium~2.0 suite comprises three libraries: \texttt{bigraph-schema} (for typed schemas and states), \texttt{process-bigraph} (for simulation interfaces and runtime), and \texttt{viva-superpowers} (for AI-assisted construction and composition of process bigraphs).

Throughout the paper, we distinguish (i) a \emph{process bigraph} as formal, typed computational object, (ii) the \emph{Process Bigraph framework} as the conceptual composition protocol, and (iii) the \texttt{process-bigraph} library as the reference implementation in Vivarium~2.0.
This article focuses on the protocol, conceptual tools, and graphical visualizations, while the Supplement details the formal machinery and provides links and descriptions of the Vivarium~2.0 software repositories.

\begin{figure}[!h]
\centering
\includegraphics[width=\linewidth]{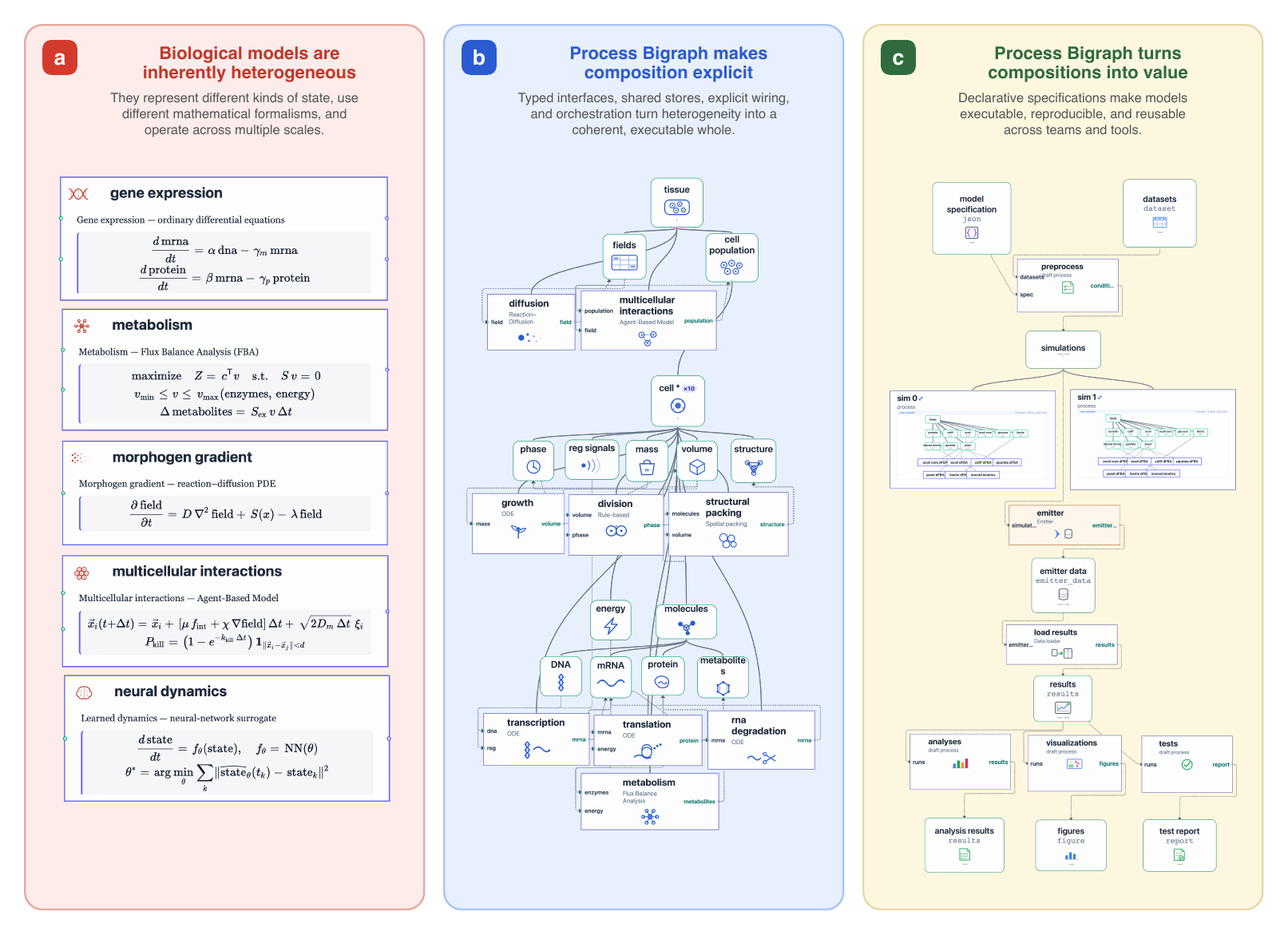}
\caption{\textbf{Process Bigraph enables heterogeneous biological models to be composed into executable, reusable multiscale simulations.}
\textbf{(a)} Biological processes are modeled using diverse mathematical formalisms, each with different variables, assumptions, and scales.
\textbf{(b)} Process interfaces provide a common basis for composition by exposing typed variables and making coupling between models explicit. This allows independently developed models to be connected across formalisms and scales.
\textbf{(c)} Process Bigraph encodes these components, interfaces, and couplings as declarative specifications that can be executed, analyzed, tested, and shared through a common engine. Standardized processes and schemas can then be reused and recomposed, providing a foundation for open-ended, community-built multiscale models.
All diagrams in this paper are drawn from real, declared process bigraphs, rendered with vivarium-workbench.}
\label{fig:overview}
\end{figure}

Figure~\ref{fig:overview} summarizes how Process Bigraph enables heterogeneous biological models to be composed into executable, reusable multiscale simulations.
Biological processes are described using diverse mathematical formalisms, each with its own variables, assumptions, and scales (Fig~\ref{fig:overview}a).
Process interfaces provide a common basis for composition by exposing typed variables and making the coupling between models explicit, allowing independently developed models to be connected across formalisms and scales (Fig~\ref{fig:overview}b).
Process Bigraph then encodes these components, interfaces, and couplings as declarative specifications that can be executed, analyzed, tested, and shared through a common engine, so that standardized processes and schemas can be reused and recomposed into a foundation for open-ended, community-built multiscale models (Fig~\ref{fig:overview}c).

We begin with the background and conceptual foundations of Process Bigraph in Section~\ref{sec:composition_framework}. 
Section~\ref{sec:process_bigraph_schema} introduces the Process Bigraph framework, including how to define process interfaces, composition patterns, and orchestration patterns, and protocols for deploying them across a heterogeneous computational infrastructure. 
Section~\ref{sec:spatioflux} introduces \emph{Spatio-Flux}, a standalone application for the multiscale simulation of spatial microbial ecosystems that demonstrates how process bigraphs support the flexible composition of diverse types and processes into complex multiscale biological simulations.
Finally, in Section \ref{sec:availability_future}, we discuss how this protocol establishes the foundations for shared infrastructure in systems biology: enabling collaborative model development, integration with existing data resources, and compositional systems biology simulations at scale.


\section{Design and Implementation}
\label{sec:design_implementation}

\subsection{Conceptual foundations}
\label{sec:composition_framework}

The process-bigraph framework is grounded in three fundamental criteria for compositional modeling and simulation.  
First, \emph{process interfaces} define the precise points of interaction between mechanisms and the system state, specifying which variables are read, written, or transformed.  
Second, \emph{composition patterns} determine how independently defined processes are coupled through shared state, enabling multiscale and multi-mechanism integration without requiring monolithic model definitions.  
Third, \emph{orchestration patterns} govern the temporal execution of processes, specifying when and how processes are invoked while ensuring consistent access to shared state.  
Together, these criteria provide an expressive foundation for constructing coherent simulations from heterogeneous components.

\begin{figure}[!h]
\centering
\includegraphics[width=\linewidth]{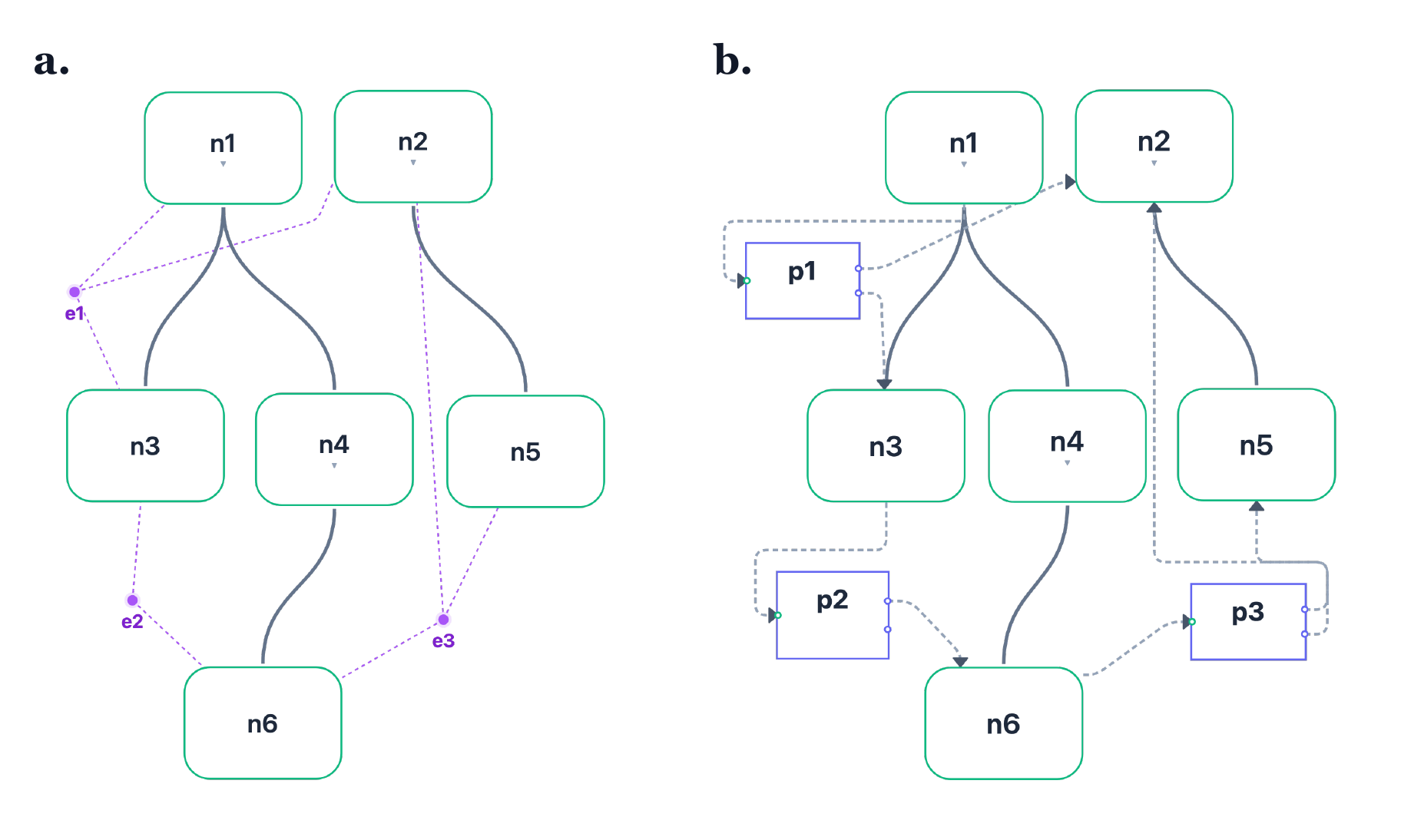}
\caption{\textbf{Composition framework overview.}
\textbf{a.} The original ``Milner'' bigraphs consist of a link graph with hyperedges shown by dashed lines ($e_1, e_2, e_3$) and nodes as rounded rectangles ($n_1, n_2, n_3, n_4, n_5, n_6$), and a place graph of the solid edges connecting the nodes.
\textbf{b.}  Process bigraphs replace the link graph with a process graph, made of processes  ($p_1, p_2, p_3$) connecting to the nodes via their ports. }
\label{fig:composition_framework}
\end{figure}

This framework builds on Robin Milner’s bigraphs \cite{milner2009space} (Fig.~\ref{fig:composition_framework}a), which unifies two complementary representations of system structure: a \emph{place graph}, capturing hierarchical containment (e.g., molecules within compartments or cells within tissues), and a \emph{link graph}, capturing patterns of interaction and connectivity via hyperedges (e.g., signaling, communication, or other forms of causal coupling).  
Although developed in computer science, bigraphs offer a powerful abstraction for biological systems whose functions depend on both spatial hierarchical organization and interactions.  
The Process Bigraph framework adapts and generalizes this dual representation by retaining the hierarchical structure of place graphs while replacing the link graph with a process graph (Fig.~\ref{fig:composition_framework}b).  
This shift emphasizes dynamics and causation: rather than encoding interactions implicitly as links, processes are treated as first-class entities that are themselves conditioned on, and act upon, the evolving system.

Processes serve as the operational units of integrative simulations.  
Each process encapsulates a hypothesized mechanism --- formally, a function that reads a specified substructure of the global system state and produces updates to that state through a well-defined interface.  
By enforcing standardized interfaces, processes can represent a wide range of modeling approaches and computational roles, including numerical simulators configured by parameters or model files, cell types or tissue agents with defined phenotypes, data transformation pipelines, machine-learning models performing inference, or analysis and visualization routines.  
This process-centric abstraction supports methodological heterogeneity while preserving composability, allowing independently developed mechanisms to be assembled into unified simulations.

The orchestration of processes draws on principles from discrete-event co-simulation, particularly the Discrete Event System Specification (DEVS) formalism \cite{zeigler2000theory}.  
In this approach, processes are scheduled as events that operate on a shared state, enabling coordination between continuous-time dynamics and discrete updates.  
The orchestration layer determines execution order, time advancement, and synchronization, while ensuring that all processes observe a consistent system state.  
This approach provides a principled mechanism for coupling processes with differing time scales, update rules, and numerical representations within a single executable model.

In Vivarium~1.0, the compositional logic of process bigraphs was implemented directly in Python code. 
While this design enabled rapid development and execution, it tightly coupled the conceptual structure to a specific implementation, limiting transparency, reuse, and interoperability.  
As models grew in complexity and demands for reproducibility, portability, and user accessibility increased, these embedded patterns proved insufficient.  
This motivated the need to make the specification of process bigraphs explicit and implementation-independent, providing a more general foundation for open-ended model construction and cross-system compatibility.

\subsection{Process Bigraph framework}
\label{sec:process_bigraph_schema}

The Process Bigraph framework provides a principled framework for composing processes, hierarchical object stores, and orchestration patterns into executable multiscale models.
Their organization aligns closely with biological semantics: processes correspond to mechanisms that drive change, hierarchy represents locations and the systems they contain, and orchestration patterns capture how heterogeneous mechanisms interact across scales in mutually constraining ways.
This correspondence becomes explicit in the Spatio-Flux example (Section~\ref{sec:spatioflux}), where metabolic fluxes, chemical fields, and particle properties derive functional roles from their positions within typed stores and interfaces.
In this section, we focus on the conceptual and computational semantics of the framework, introducing its core abstractions through intuition, diagrams, and examples, while the Supplement 1 presents formal definitions.

Table~\ref{box:compositional} summarizes the core vocabulary of the framework, linking each conceptual term to its formal anchor and definition.
Together, these elements define a language for expressing how heterogeneous processes can be connected and orchestrated across scales.
Paths, types, and values ($P, T, V$) provide the basic addressing and data structures, while schemas and states ($\Sigma, x$) define the typed organization of the system.
Processes ($f$) interact with shared state through ports and wiring ($W$), emitting typed updates ($\delta$) that are applied according to type-specific rules ($apply_\tau$) defined in the type registry ($R_T$).
The link registry ($R_L$) connects abstract process definitions to executable implementations, while orchestration specifies how the resulting system evolves over time.
Supplement 1 details the type system, composition rules, orchestration methods, and update semantics used during execution, while this section develops their conceptual interpretation.

A central principle of Process Bigraphs is the separation between \emph{schema} ($\Sigma$) and \emph{state} ($x$).
The schema defines the typed organization of the hierarchy, while the state assigns concrete values to those paths for a particular simulation run.
Together, $\Sigma$ and $x$ define a process bigraph $B = (\Sigma, x, R_T, R_L)$, in which typed stores function as shared variables and processes operate through explicitly wired interfaces.
This separation enables model validation, structural reuse, and reasoning about composition independently of execution.
The type system ensures compatibility between connected components and defines how concurrent updates to shared variables are combined.

\begin{table}[h]
\centering
\small
\setlength{\tabcolsep}{6pt}
\renewcommand{\arraystretch}{1.18}
\begin{tabularx}{\linewidth}{>{\raggedright\arraybackslash}p{0.18\linewidth} >{\raggedright\arraybackslash}p{0.23\linewidth} X}
\toprule
\textbf{Term} & \textbf{Formal anchor} & \textbf{Description} \\
\midrule
Paths, types, and values & $P$ paths, $T$ types, and $V$ values & These are the ingredients needed to name locations in a hierarchical system, declare what may be stored there, and assign concrete values. \\

Schema & $\Sigma : P \rightharpoonup T$ & The schema is the typed organizational blueprint of the system. \\

State and stores & $x : P \rightharpoonup V$ & The state assigns values to schema paths, and stores are the state locations that function as shared variables. \\

Process & $f : (in_{\tau_1}, \dots) \rightarrow (out_{\tau_2}, \dots)$ & A process is a mechanism with an interface, defined by what it reads, what it can update, and its configuration. \\

Deltas and typed application & $\delta = update(x_{\mathrm{in}}, \Delta t)$ and $x' = apply_{\tau}(x,\delta)$ & Processes emit typed updates rather than directly overwriting state, and each type determines how its updates are applied. \\

Ports and wiring & $W : (\text{process}, \text{port}) \rightarrow P$ & Ports specify accessible variables, and wiring makes every coupling explicit by mapping each port to the state path it reads or writes. \\

Type registry & $R_T$ & The type registry defines available types and their validation and update semantics, including the operator $apply_{\tau}$. \\

Link registry & $R_L$ & The link registry maps process addresses to concrete handler classes, connecting abstract process nodes to executable implementations. \\

Process bigraph & $B = (\Sigma, x, R_T, R_L)$ & A process bigraph is the typed and executable organization as a whole, with its wiring pattern derived from the port bindings stored in the state. \\

Composite and bridge & $bridge : (\text{external port}) \rightarrow P$ & A composite packages an internal process bigraph as a single higher-level process and maps its external interface onto internal state paths. \\

Orchestration & $\langle B, t, t_{next} \rangle \rightarrow \langle B', t', t'_{next} \rangle$ & Orchestration specifies when processes run, how their deltas are combined, and how structural rewrites are incorporated over time. \\
\bottomrule
\end{tabularx}
\caption{Core vocabulary of Process Bigraph. Detailed definitions are provided in the Supplementary Materials.}
\label{box:compositional}
\end{table}

This structure is developed in the remainder of the section by elaborating the core elements introduced in Table~\ref{box:compositional}.
Types and stores (subsection~\ref{sec:types-stores}) provide the shared semantic backbone that enables independent processes to interoperate.
Interfaces (subsection~\ref{sec:process-interface}) define the points of contact through which processes can be composed, substituted, or refined.
Composition patterns (subsection~\ref{sec:composition-patterns}) describe how processes coordinate through shared stores to produce coherent multilevel behavior.
Orchestration patterns (subsection~\ref{sec:orchestration}) determine how these connected processes unfold over time, shaping and being shaped by the states they act upon.

\subsection{Types and stores}
\label{sec:types-stores}

Given a schema $\Sigma : P \rightharpoonup T$, the state $x : P \rightharpoonup V$ assigns concrete typed values to paths in the hierarchy.

While schemas describe the structure of a composite model, types describe the semantic meaning of the data that flows through it.
They capture the expected form, units, and constraints of the data, so that processes can exchange information consistently even when they come from different modeling perspectives or operate at different scales.

Stores correspond to state locations that processes may access through ports and wiring relations.

\begin{figure}[!h]
\centering
\includegraphics[width=\linewidth]{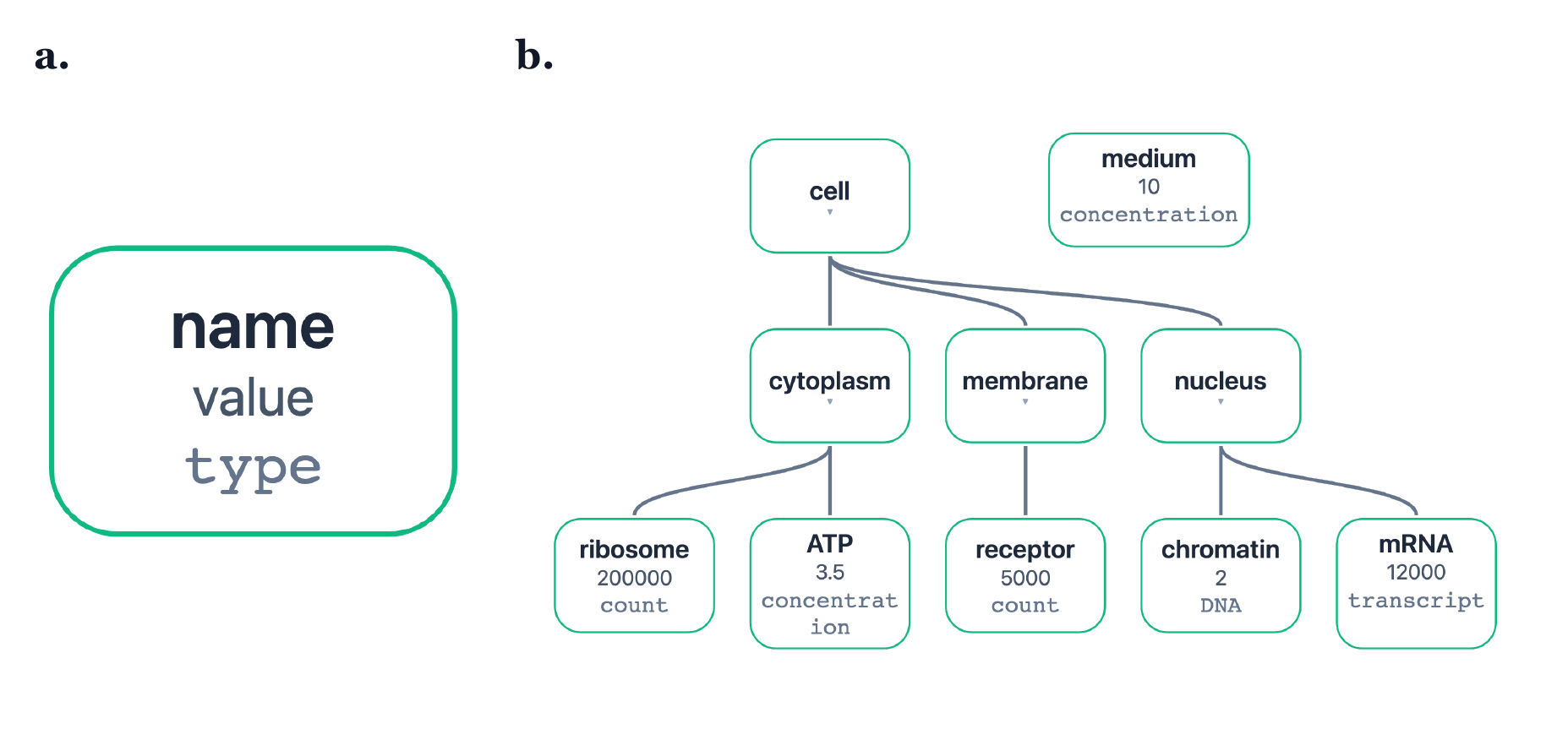}
\caption{\textbf{Store diagrams.}
\textbf{a.} A store is depicted as a rounded rectangle and can hold any data type; it displays its name, value, and type.
\textbf{b.} Stores nest in hierarchies for multiscale representation: a place graph of nested stores shows outer stores connected to inner stores by edges, with each store showing its name, value, and type (e.g., a cell containing cytoplasm, membrane, and nucleus).}
\label{fig:store}
\end{figure}

Stores (Fig.~\ref{fig:store}) hold the simulation state.  
Each store is associated with a type, which determines what processes may read from or write to it and how updates to that store should be interpreted.
Types may range from simple primitives (numbers, arrays, strings) to richer scientific objects such as molecular concentrations, reaction rates, spatial fields, or images with coordinate metadata.
The type labels shown in Figs.~\ref{fig:store}--\ref{fig:composite} (e.g., ``concentration'', ``species'', ``DNA'') are intentionally informal and illustrative; Supplement~1 introduces the more precise schema-level type definitions and update semantics used by the framework.
By assigning explicit types to stores, process bigraphs make the data dependencies between models visible and checkable: a process can be rewired or substituted only if its port types remain compatible with the attached stores.

When stores are nested, as branches in a place graph (Fig.~\ref{fig:store}b), they can encode hierarchical biological organization --- for example, molecules inside organelles, organelles inside cells, and cells inside tissues.
This hierarchical structure provides a substrate for multiscale modeling: local processes act on specific stores, while other processes aggregate, coarse-grain, or constrain behavior across other parts of the tree.

\subsection{Process interface}
\label{sec:process-interface}

Processes are formally defined by typed interfaces $f : (in_{\tau_1}, \dots) \rightarrow (out_{\tau_2}, \dots)$ together with configuration state and update behavior.

Each process participates in a process bigraph by exposing a clear interface: a set of input ports, output ports, and expected configuration parameters (``config'').
Conceptually, a process is a mechanism that observes part of the current state, uses its configuration to decide how to respond, and proposes changes to the state through its outputs --- much like the concept of a function both in mathematics and in biology.

\begin{figure}
\centering
\includegraphics[width=0.7\linewidth]{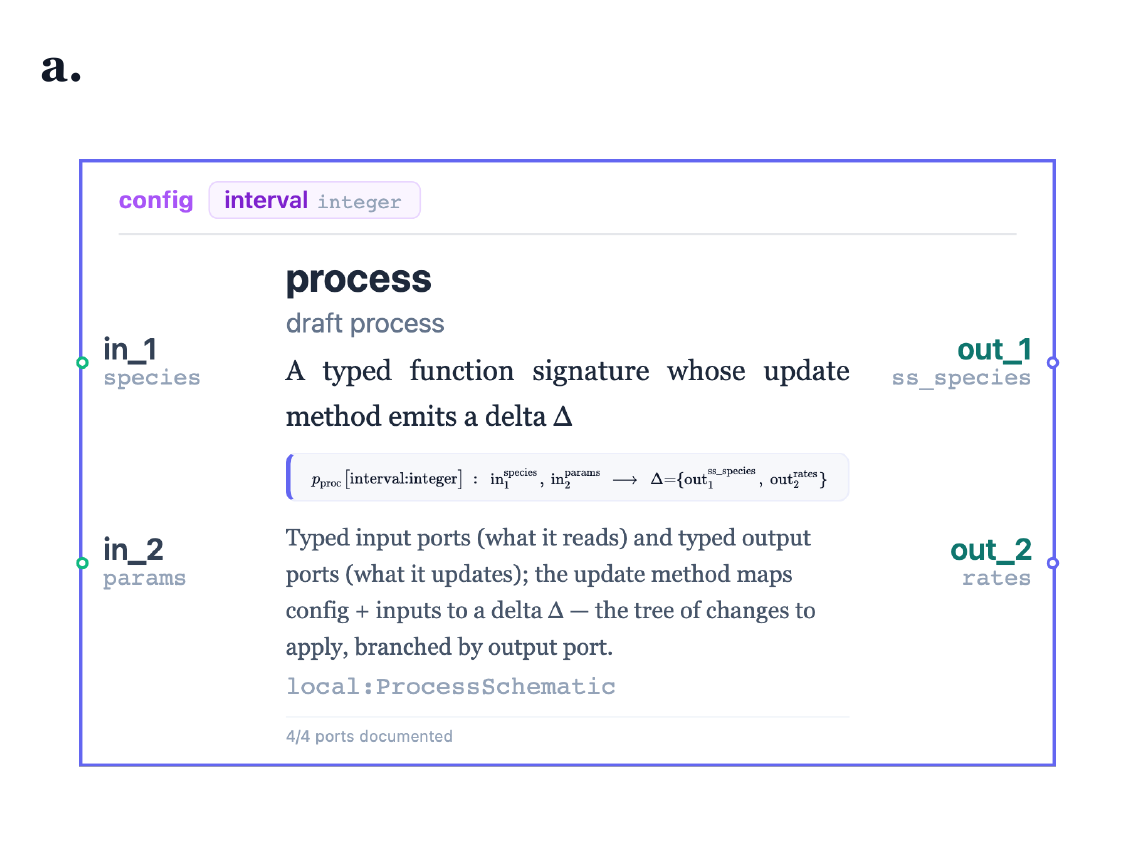}
\caption{\textbf{Process diagram.}
A process is depicted as a rectangle with ports along its boundary to represent its interface.
Input ports are here shown on the left and outputs on the right.
Ports specify the type of information flowing in and out (``species'', ``params'' as inputs; ``ss\_species'', ``rates'' as outputs).
The process's update function maps its inputs to a typed delta of outputs, informed by its config (here, an ``interval'' parameter of type integer).}
\label{fig:process}
\end{figure}

As illustrated in Fig.~\ref{fig:process}, a process is drawn as a rectangle with input and output ports along its boundary.
Ports advertise the type of data that flows through them, to ensure that information is passed consistently between processes.
Configuration values capture anything that is fixed for a given instance of the process: model parameters, file paths, learned weights, or other metadata.

Ports provide constrained access to shared stores, separating a process implementation from the state paths to which it is wired.

This interface-driven perspective decouples internal implementation details.
A process might wrap a differential equation solver, a flux balance analysis model, a rule-based simulator, a neural network, a data transformation, or a visualization routine.
From the perspective of the process bigraph, all of these are just processes with typed inputs, outputs, and config.
This abstraction allows different modeling methods to coexist within a single simulation while remaining swappable and reusable.

\subsection{Composition patterns}
\label{sec:composition-patterns}

Composition patterns arise from the wiring relation $W : (\text{process}, \text{port}) \rightarrow P$, which determines how process interfaces attach to shared stores.

\begin{figure}[!h]
\centering
\includegraphics[width=\linewidth]{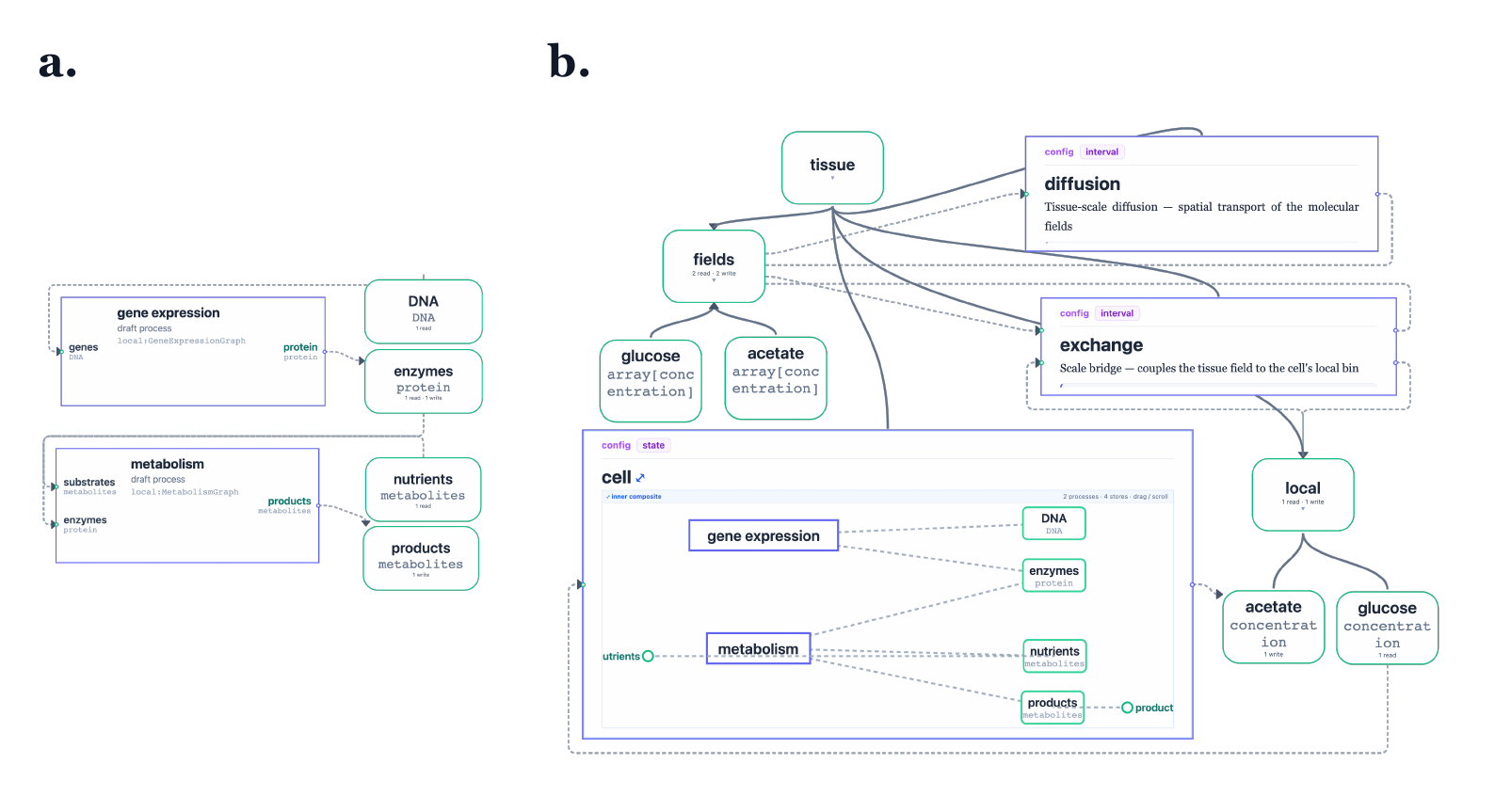}
\caption{\textbf{Composite diagrams.}
\textbf{a.} A process graph consists of processes (i.e., \textit{metabolism}, \textit{gene expression}) connected to stores (i.e., \textit{DNA}, \textit{enzymes}, \textit{nutrients}, \textit{products}) via ports (i.e., genes, substrates, products, etc.).
Port types must match the connected store type (i.e., the enzymes port must connect to an enzyme type store), with inputs/outputs indicated by arrow directions. 
\textbf{b.} A composite process contains an internal process bigraph that it orchestrates. 
The composite has external input/output ports that can connect to external stores, with matching internal ports linked to the internal bigraph via dotted wires.
Variables connected across these inner/outer port pairs can pass updates to remain synchronized.}
\label{fig:composite}
\end{figure}

A composition pattern specifies how processes are connected to stores --- and through shared stores to each other.
At a single level without hierarchy, these patterns define a process graph (Fig.~\ref{fig:composite}a) in which multiple processes interact through shared stores.
When nesting is introduced, processes may connect across levels, enabling communication between micro-scale and macro-scale behavior.

The process bigraph emerges from the superposition of the place graph hierarchy and the process graph wiring structure.

The combined structure of the place graph (the hierarchy of stores) and the process graph (the wiring between ports and stores), along with its corresponding schemas and types, constitutes a process bigraph (Fig.~\ref{fig:composition_framework}b).  
Composition patterns express modeling choices that rarely appear explicitly in multiscale simulation: which processes share which variables, which scales interact directly, and how information flows between subsystems.

A composite is a model built from other models: it bundles an internal process bigraph together with an external interface.
To connect internal and external state, the composite exposes \emph{bridge} ports that link external stores to corresponding stores in the internal place graph (Fig.~\ref{fig:composite}b).
These bridges synchronize data across the composite boundary and allow the composite to be nested inside higher-order composites just like any other process.

Because composites behave just like ordinary processes, entire hybrid simulations can be encapsulated behind an interface and then assembled into larger ``super-simulations.''
From the process-bigraph viewpoint, this is equivalent to zooming into a composite to reveal its internal structure, or replacing an atomic process with a more detailed composite that exposes the same interface.
Such substitutability provides a natural mechanism for fine- and coarse-graining strategies in multiscale modeling.

\subsection{Orchestration patterns}
\label{sec:orchestration}

While composition determines structural connectivity, orchestration determines how process-generated deltas are scheduled, merged, and applied over time.

Composition patterns describe how processes are wired; orchestration patterns describe how they are run in time.
Composites serve as orchestrators that govern the execution of their internal processes: at each orchestration step, the composite decides which processes are eligible to run, gathers the information they need from the state, invokes them, and incorporates their proposed changes back into the shared stores.

Operationally, processes emit typed updates $\delta = update(x_{\mathrm{in}}, \Delta t)$ that are incorporated into the shared state through type-specific application rules $apply_\tau$.

\begin{figure}[!h]
\centering
\includegraphics[width=\linewidth]{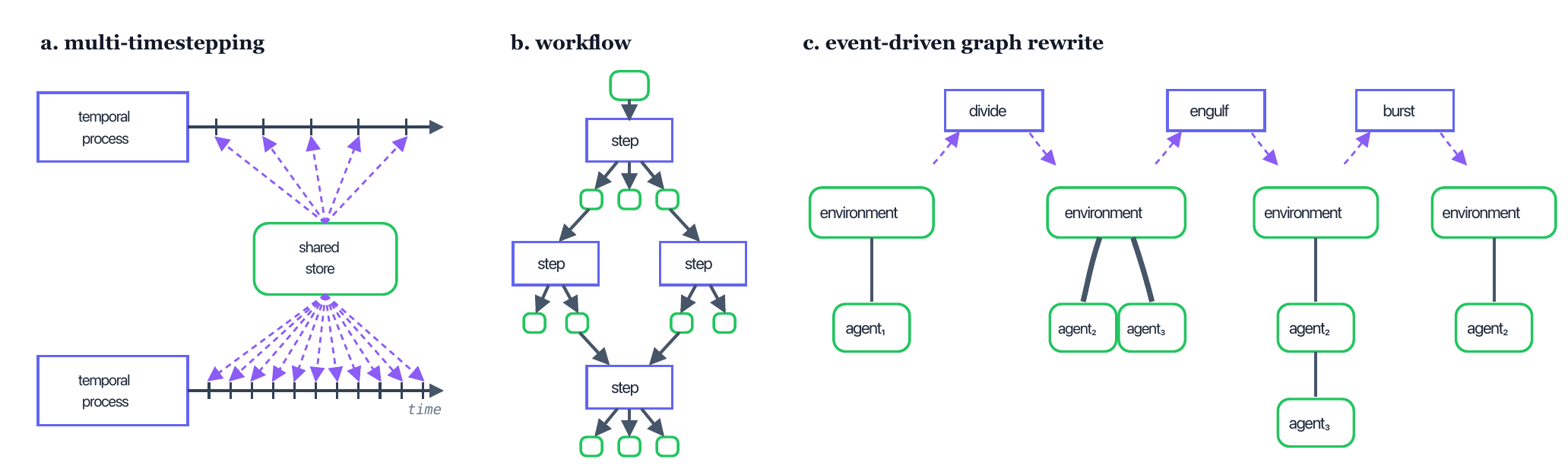}
\caption{\textbf{Orchestration patterns.}
\textbf{a.} Multi-timestepping, with temporal processes each updating at their preferred time intervals, orchestrated by a discrete event co-simulation engine. 
\textbf{b.} A Workflow is a directed acyclic graph that sets the order of updates for step processes, with each one triggered by changes to its input dependencies. 
\textbf{c.} Event-driven graph re-write processes orchestrated as discrete events, which change the topological structure of an agent-environment system.
Each graph re-write can be specified as a reaction.
``Divide'' makes one agent divide into two, ``engulf'' place one agent inside of another, ``burst'' dissolves the agent, releasing its inner components back into its environment.}
\label{fig:orchestration_diagram}
\end{figure}

Figure~\ref{fig:orchestration_diagram} summarizes three orchestration patterns available for Vivarium~2.0:
(1) Multi-timestepping (Fig.~\ref{fig:orchestration_diagram}a) coordinates continuous and discrete processes that evolve on different timescales, using discrete-event simulation adapted from the Discrete Event System Specification formalism \cite{zeigler2000theory,goldstein2018multiscale}.  
Each temporal process declares its preferred update interval, and the orchestrator advances time by triggering whichever process is scheduled to run next, while keeping all processes coupled through the shared state.
(2) Workflows (Fig.~\ref{fig:orchestration_diagram}b) arrange processes in directed acyclic graphs, so that each step runs only after its dependencies have produced or updated their outputs.  
This is useful for initialization routines, simulation experiment workflows, analysis pipelines, or any computation with a natural order.
(3) Event-driven structural updates (Fig.~\ref{fig:orchestration_diagram}c) allow the composite to change its own structure.  
Rules such as division, engulfment, or bursting can insert, remove, or rewire processes and stores in response to state-dependent conditions.
In all cases, the process updates arrive as typed \emph{updates} that must be merged into the shared state, with conflicts managed by the type engine.

These patterns are interoperable: workflows can sit between temporal events, structural updates can be triggered either by scheduled times or internal state, and entire composites can themselves be embedded as processes within larger composites.
Together, these can be used to describe how rich multiscale behavior unfolds, without tying the framework to any single numerical method or simulator.

\subsection{Execution Protocols}
\label{sec:protocol}

A process interface protocol specifies the minimal behavior a process must exhibit in order to participate in a process-bigraph composition.
This requires that a process declares the types of data it reads and writes through its ports, accepts a configuration, and exposes an update method that proposes changes to the shared stores.

The protocol is agnostic to how a process is implemented or where it runs -- it might be a local Python function, a compiled simulator, a containerized service, or a remote model accessed over a network.
As long as it presents a compatible interface and communicates using the agreed schema and state representation, it can be wired into a process bigraph and orchestrated alongside other components.
In the Supplement we describe several concrete execution protocols that realize this idea: local in-memory execution, parallel execution, container-based simulators, and REST-style services.
The key point is that a shared interface protocol allows diverse tools and models to coexist within a common architectural layer, making it possible to build complex simulations from independently developed pieces.

\section{Results}
\label{sec:results}

\subsection{Spatio-Flux: a worked example of process bigraphs}
\label{sec:spatioflux}

We constructed a series of composite simulations in \emph{Spatio-Flux}, a stand-alone library developed to demonstrate the composition of metabolic, spatial, mechanical, and structural processes within a single process bigraph type system.
Spatio-Flux was designed as a testbed for composing independently developed mechanisms through typed interfaces, rather than as an optimized or quantitatively calibrated domain-specific simulator.
Parameter values across all examples were chosen primarily at the level of order-of-magnitude estimates that yield qualitatively interesting and dynamically nontrivial behavior, rather than to match any specific experimental system.
The goal of these simulations is to illustrate compositional structure and cross-process interaction, not to produce predictive biological models, while providing a foundation that can be systematically refined toward predictive use as additional data, constraints, and calibrated components are incorporated.
To test whether the same compositional implementation can reproduce an established spatial dFBA reference, Supplement~2 includes a matched comparison between Spatio-Flux and the COMETS Java backend through \texttt{cometspy} \cite{dukowski2021comets}
That comparison shows that Spatio-Flux reproduces COMETS biomass trajectories and spatial metabolite fields within numerical tolerance when geometry, kinetics, diffusion coefficients, and initial conditions are matched.

\begin{figure}[H]
\centering
\includegraphics[width=\linewidth]{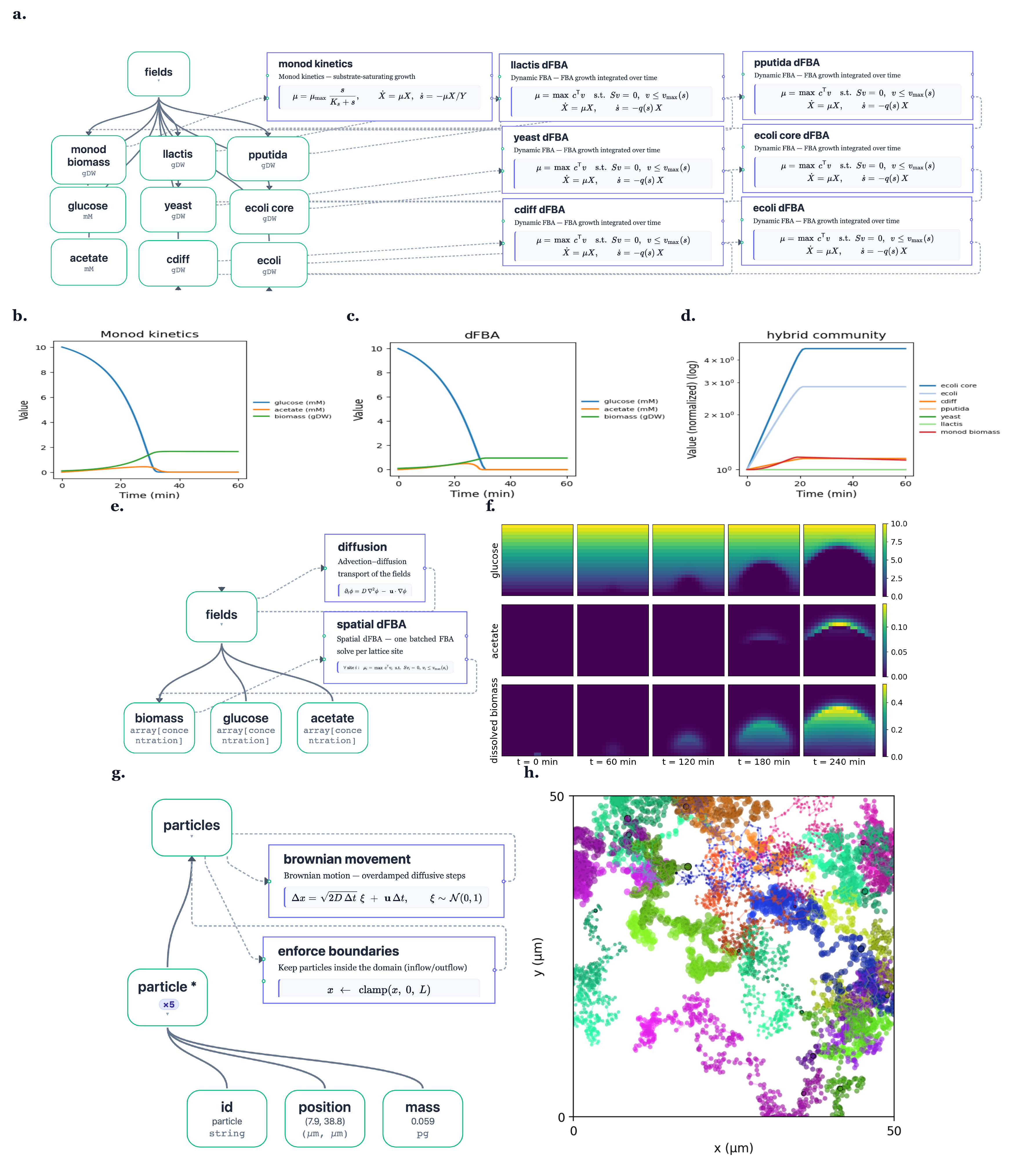}
\caption{\textbf{Spatio-Flux basic components and behaviors.}
\textbf{a.} Community dFBA dynamics for multiple microbial species, including \emph{E.~coli} (core and iAF1260), \emph{C.~difficile}, \emph{P.~putida}, \emph{S.~cerevisiae}, and \emph{L.~lactis}, all interacting through shared nutrient fields. There is also a single Monod kinetic process in the community.
\textbf{b.} Single-species growth driven by Monod kinetics.
\textbf{c.} Single-species growth governed by dynamic flux balance analysis (dFBA).
\textbf{d.} Results of the community model depicted in panel a.
\textbf{e.} A reconstructed COMETS simulation, with dFBA processes distributed across a lattice and metabolites transported through diffusion.
\textbf{f.} Results for the COMETS simulation, with snapshots of spatial fields showing the evolution of nutrient concentrations and biomass over time.
\textbf{g.} Brownian particles -- a particle-based simulation in which particles undergo stochastic diffusion via a Brownian movement process, with positions updated in continuous space and boundary conditions enforced by a separate \texttt{enforce\_boundaries} process.
\textbf{h.} Traces of particle trajectories over the course of the Brownian particle simulation, illustrating their spatial exploration of the domain.}
\label{fig:multi_dfba_viz}
\end{figure}

\begin{figure}[H]
\centering
\includegraphics[width=\linewidth]{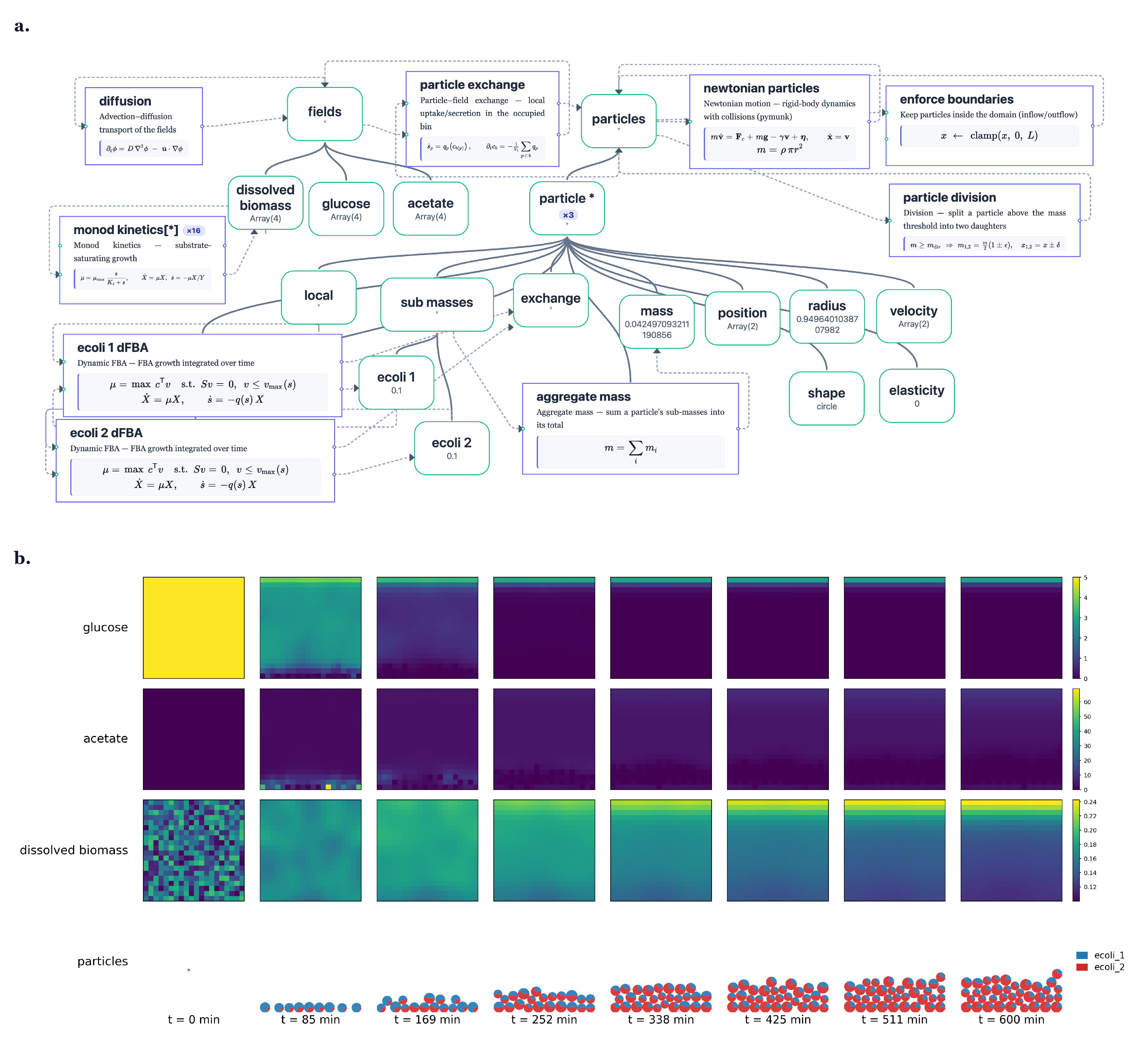}
\caption{\textbf{Spatio-Flux reference model.}
\textbf{a.} The integrated composite simulation, showing how metabolic, spatial, mechanical, and structural processes are connected through shared typed stores.
Newtonian particles carry internal metabolic processes and exchange metabolites with a spatial lattice via particle--field adapters, while diffusion updates field concentrations and graph-rewrite processes govern particle division and boundary interactions.
\textbf{b.} Representative simulation snapshots from this composite, showing particles moving through and reshaping spatial nutrient fields.
Three fields are shown: glucose, acetate, and dissolved biomass which is using a kinetic model.
The particles are also interacting with the glucose and the acetate fields, through uptake and secretion.
On the particles there are two species: ecoli\_1, which prefers glucose, and ecoli\_2, which prefers acetate -- their relative proportions change as the system evolves and available fields change.}
\label{fig:spatioflux_metacomposite}
\end{figure}

Figure~\ref{fig:multi_dfba_viz} summarizes a range of simpler composites, from single-site metabolic models to spatially extended, particle-based community simulations, while Fig.~\ref{fig:spatioflux_metacomposite} presents a larger combined reference simulation integrating multiple interacting processes within a microbial ecosystem.
This section describes a representative subset of possible composites, using the larger composite as a unifying reference.
Detailed descriptions of the processes and parameters are provided in Supplement 2.
Supplement~2 also reports runtime decomposition and wall-clock scaling for these examples, including local and distributed execution modes.
The online Spatio-Flux repository includes a public automated test suite with many additional composites, including mixed Monod--dFBA communities, spatial lattices, Brownian and Newtonian particle systems, and integrated particle--field simulations; this test suite is intended as a flexible and evolving medium, and individual example configurations may change over time as the library and its capabilities expand.

All Spatio-Flux simulations are assembled from processes with explicit state types, input--output interfaces, and update rules.
State is organized into hierarchically nested stores that represent biologically meaningful structure, such as environments containing spatial fields and particles containing internal metabolic state.
Processes are composed without modification when their interfaces are compatible, allowing models developed independently to be recombined within a single simulation.
The examples shown here draw from five process families: local metabolic species, field movement, particle movement, particle--field adapters, and graph-rewrite processes (Table~\ref{tab:spatioflux-processes}).

At the simplest level, single-site metabolic simulations were constructed using either Monod kinetics \cite{monod2012growth} or dynamic flux balance analysis (dFBA).
Both formalisms operate on the same \texttt{substrates} store, which represents extracellular metabolite concentrations, and expose identical input--output types describing substrate uptake, secretion, and biomass change.
Parameters for these models were selected to ensure stable growth and depletion dynamics over accessible simulation timescales, rather than to reflect organism-specific kinetic measurements.
Figures~\ref{fig:multi_dfba_viz}b and c show the resulting Monod and dFBA trajectories.
Because their interfaces match, these processes are substituted without altering surrounding components, enabling hybrid compositions in which different metabolic formalisms coexist within a shared environment (Fig.~\ref{fig:multi_dfba_viz}d).

Community-level simulations were constructed by placing multiple local metabolic species into a shared nutrient environment.
Each species specifies its own metabolic model, exchange reactions, and parameters, while all metabolic processes bind to the same environmental stores.
Metabolic models were selected at random from the BiGG database and paired with generic uptake bounds and kinetic parameters chosen to produce heterogeneous but comparable growth behaviors.
These models are not intended to represent a realistic or curated co-culture; example species include \emph{E.~coli}, \emph{S.~cerevisiae}, and \emph{L.~lactis}.
To further demonstrate compositional flexibility, a single Monod kinetic process was included alongside the dFBA models within the same environment.
Species interact exclusively through typed environmental fields, with each consuming and secreting substrates according to its results while the environment reflects the combined effects of all species.
This interaction structure, in which coupling is mediated entirely through shared state rather than direct process-to-process communication, is shown schematically in Fig.~\ref{fig:multi_dfba_viz}a.
The resulting timeseries in Fig.~\ref{fig:multi_dfba_viz}d show the different species growing until resources are depleted.

Spatially explicit simulations were produced by introducing field-movement processes that update nutrient fields through diffusion.
Diffusion coefficients, uptake rates, and lattice resolutions were chosen to generate visible gradients and spatially structured growth within short simulation runs, rather than to reproduce measured transport properties.
In these composites, each lattice site hosts a local metabolic species, and neighboring sites exchange substrates via finite-volume diffusion processes.
Metabolic uptake and secretion occur locally at each site, while diffusion propagates gradients across the lattice.
This construction reproduces the core algorithmic structure of COMETS-style spatial dFBA, in which metabolism and transport are alternated on a lattice to produce spatiotemporal metabolic dynamics \cite{dukovski2021metabolic}.
In Supplement~2, we use this correspondence to perform a direct COMETS comparison under matched assumptions.
For the matched $N=16$ benchmark, Spatio-Flux and COMETS agree on total biomass within $0.1\%$, with close agreement in glucose, acetate, and biomass spatial fields.
Figures~\ref{fig:multi_dfba_viz}e--f show the resulting spatiotemporal dynamics, with a gradient of non-diffusing glucose and diffusing biomass and acetate.
As a result, biomass moves up the glucose gradient and grows more rapidly as it proceeds, while acetate released by growth is subsequently consumed.

Particle-based simulations were constructed by introducing motile particles governed by alternative particle movement processes.
A basic \texttt{BrownianMovement} process implements stochastic motion in a continuous spatial domain.
The \texttt{PymunkParticleMovement} process extends this representation with Newtonian dynamics, adding attributes such as mass, velocity, and elastic interactions through the \texttt{pymunk} physics engine.
Physical parameters such as masses, forces, and damping were selected to produce qualitatively distinct regimes of motion and interaction, rather than to correspond to specific biophysical measurements.
Figures~\ref{fig:multi_dfba_viz}g--h illustrate simulations with the Brownian movement process, showing particles diffusing through continuous space.
Population structure is updated through graph-rewrite processes that manage boundary interactions, adding or removing particles in response to state-dependent conditions and timestep-dependent rates.

Figure~\ref{fig:spatioflux_metacomposite} presents the large reference simulation that integrates all previously described process families into a single executable model of a microbial ecosystem.
This simulation includes both dissolved biomass represented in spatial fields and particle-associated biomass carried by motile particles.
In this composite, each particle carries two distinct internal dFBA models derived from modified \emph{E.~coli} core metabolism.
One internal model is optimized for glucose uptake (ecoli1), while the other is optimized for acetate uptake (ecoli2).
Exchange bounds and objective weights were selected to ensure that each model dominates under different substrate regimes, enabling visible metabolic switching rather than quantitatively accurate pathway usage.
Particle motion is governed by Newtonian dynamics, introducing gravity and mechanically mediated interactions such as collisions, crowding, and spatial segregation.
The surrounding environment is represented as a spatial lattice governed by Monod kinetics and diffusion, producing continuously evolving chemical fields.
Particle--field exchange processes connect continuous particle positions to discrete environmental fields, allowing particles to sample local concentrations and apply metabolic exchanges.
An accumulate-mass adapter aggregates biomass production from all internal metabolic processes into a single particle-level mass.
A graph-rewrite process monitors this total mass and triggers division when a threshold is exceeded, restructuring the population in response to emergent growth dynamics.
Together, these metabolic, spatial, mechanical, and structural processes are composed through typed interfaces and coordinated within a shared orchestration framework.
The resulting dynamics, shown in Fig.~\ref{fig:spatioflux_metacomposite}b, illustrate a qualitative succession in which glucose-specialist populations dominate early, followed by acetate-specialist populations as byproducts accumulate.

\begin{table}[ht]
  \centering
  \renewcommand{\arraystretch}{1.25}
  \begin{tabular}{p{0.2\textwidth} p{0.26\textwidth} p{0.46\textwidth}}
    \hline
    \textbf{Process family} & \textbf{Processes} & \textbf{Role in composite simulations} \\
    \hline

    Metabolic processes &
    \texttt{DynamicFBA}, \texttt{MonodKinetics}, \texttt{SpatialDFBA} &
    Compute metabolic uptake, secretion, and biomass production at single sites or across spatially distributed locations by operating on substrate fields and biomass variables. \\

    Field transport &
    \texttt{DiffusionAdvection} &
    Updates dissolved species fields through diffusion and advection, coupling metabolic activity across space. \\

    Particle movement &
    \texttt{BrownianMovement}, \texttt{PymunkParticleMovement} &
    Updates particle positions in continuous space. Brownian motion provides stochastic movement, while Newtonian motion introduces mass, velocity, inertia, friction, and elastic interactions. \\

    Particle--field coupling &
    \texttt{ParticleExchange} &
    Mediates bidirectional exchange between particle-local state and spatial fields, synchronizing internal particle chemistry with nearby lattice values. \\

    Structural and boundary processes &
    \texttt{ParticleDivision}, \texttt{ManageBoundaries} &
    Rewrites the particle store by creating, removing, or relocating particles in response to state-dependent conditions such as growth thresholds or boundary crossings. \\

    \hline
  \end{tabular}
  \caption{    \textbf{Spatio-Flux process families.}
    Each family encapsulates a distinct modeling concern—metabolism, transport, motion, exchange, or structural change—that can be composed through typed interfaces.
    More complex simulations arise by combining families rather than extending any single process.}
  \label{tab:spatioflux-processes}
\end{table}

\section{Availability and Future Directions}
\label{sec:availability_future}

\subsection{Software availability}
Vivarium 2.0 and the software described in this article are openly available through the Vivarium Collective GitHub organization at \url{https://github.com/vivarium-collective}.
The principal libraries are \texttt{bigraph-schema}, \texttt{process-bigraph}, and \texttt{viva-superpowers}; Spatio-Flux is provided as a reference application and reproducible demonstration suite.
A collection of reusable emitters---observational step processes that record and export simulation results to different backends without affecting simulation state---is provided in the \texttt{viva-emitters} repository.
The repositories include source code, installation instructions, usage examples, tests, and documentation for constructing and executing process-bigraph simulations.
All of the process bigraphs visualized in this paper are declared as real, executable specifications and are available to interact with in the Spatio-Flux repository, and the visualization tool used to render their diagrams, \texttt{vivarium-workbench}, is openly available in the Vivarium Collective GitHub organization.
The software is distributed under open-source licenses specified in the individual repositories.
The Supplementary Materials describe the formal framework, installation and execution details, example configurations, and reproducibility instructions for the simulations reported here.

\subsection{Future directions}

Just as the internet protocol made it possible for many different computers and services to work together, a composition protocol for systems biology can make it possible for models, datasets, and simulators developed by different groups to work together within a single simulation.
Rather than focusing on any one modeling method, such a protocol provides a shared way to connect components, exchange state, and coordinate execution over time.
This enables multiscale simulations that combine biological structure, experimental data, and mechanistic function into a single computational system, rather than treating them as separate products.

The process--bigraph protocol is designed to complement existing standards and infrastructure in computational biology by addressing how models and simulators are connected into larger systems.
Standards such as SBML and CellML describe individual biochemical and biophysical models \cite{keating2020sbml,lloyd2004cellml}, while SBML-spatial \cite{schaff2023sbml}, VCML as used by Virtual Cell \cite{loew2001virtual}, and emerging multicellular schemas \cite{fletcher2022seven} extend these descriptions to include spatial organization and agent-based structure.
Resources such as BioModels \cite{le2006biomodels} and Physiome \cite{hunter2003integration} provide curated model repositories, and platforms such as BioSimulators \cite{shaikh2022biosimulators} supply execution backends that allow these models to be run across multiple simulation engines.

Process bigraphs operate one level above individual model descriptions and execution backends.
They describe how multiple models and simulation tasks --- such as ODEs, flux balance analysis models, agent-based models, stochastic simulators, spatial solvers, and machine-learning surrogates --- are connected, how shared variables are exchanged, and how execution is coordinated.
In Spatio-Flux, this compositional layer enabled metabolic, spatial, particle-based, and mechanical processes to be substituted, recombined, and jointly executed within a single simulation, while remaining compatible with existing model formats and runtimes.
This same layer allows SBML- and CellML-based models to be executed through infrastructure such as BioSimulators following SED-ML workflows, while remaining coupled to other processes and data transformations in a multiscale context.

A practical consideration for adopting a protocol-based approach is the effort required to integrate existing simulators and models into a common interface, and to connect them.
While the process interface itself is minimal, exposing legacy tools as compatible processes still requires defining typed ports, update semantics, adapters, installation workflows, and interoperability across different environments and programming languages.
To streamline this step and make integration patterns more reproducible, we developed a GitHub repository called viva-superpowers with a set of reusable AI agent skills that scaffold process wrapping and composition.
These tools automate the creation of port-typed process interfaces and composite connection patterns, reducing manual effort and ambiguity.
At the time of writing, over a dozen simulators and models have been wrapped using this approach, including COMETS \cite{dukovski2021metabolic}, CompuCell3D \cite{swat2012compucell3d}, Mem3DG \cite{mem3dg2022}, Smoldyn \cite{andrews2010smoldyn}, VCell's finite-volume solver \cite{loew2001vcell,schaff2016vcell_spatial}, Martini \cite{marrink2007martini}, and LAMMPS \cite{plimpton1995lammps}.
These wrapped simulators can function as components within extended Vivarium simulations, while some also encapsulate internal hybrid methods.
From the perspective of process bigraphs, however, such frameworks represent only restricted classes of composites, whereas process bigraphs provide a general representation for arbitrary composite simulations.

This connective role is especially important for bringing rich biological datasets into functional simulations.
Large data resources such as HuBMAP \cite{hubmap2019human} already organize measurements into spatially nested structures, for example cells within functional tissue units and functional tissue units within tissues.
Similarly, pathway and metabolic databases such as BioCyc \cite{karp2019biocyc} and Reactome \cite{jassal2020reactome} provide structured descriptions of biological function.
Process bigraphs preserve this hierarchical organization while adding an explicit graph of typed processes that act on the data, allowing variables to acquire biological meaning through their functional role in computation.
In Spatio-Flux, this alignment is demonstrated through types that distinguish concentrations, fluxes, fields, particles, and masses, without requiring explicit linkage to external ontologies.

Typed interfaces play a practical role in making this semantic alignment sustainable.
By specifying what data each process reads and writes, they make it easier to combine independently developed components, replace one model with another, and detect incompatibilities early.
These interfaces act as lightweight contracts that can later be mapped onto established biological ontologies, rather than being hard-coded into individual simulations.
This supports long-term model maintenance, reuse, and versioning, and points naturally toward shared registries of schemas, validation tests, and community conventions.

The protocol would also strengthen reproducibility and integration with research infrastructure.
Because compositions are specified independently of execution, the same process-bigraph document can be run locally, on high-performance computing systems, or through cloud services.
In Spatio-Flux, all results in this paper are generated automatically from schemas, executed through a common runtime, and visualized through a shared pipeline.
These same compositions are exercised in an automated test suite with online GitHub-based workflows, illustrating how composite simulations can be tested, executed, and shared as part of standard computational infrastructure.

By providing a clear way to connect models, datasets, and simulators, Process Bigraph aims to support collaborative, infrastructure-driven systems biology.
Within this setting, different groups can contribute models, data products, and simulation components that remain usable outside their original context.
As these shared compositions grow, researchers will be able to integrate heterogeneous datasets directly into functional simulations, link models across biological scales, and collaboratively build increasingly predictive representations of living systems --- from microbial communities to tissues and organs --- using shared, open infrastructure.

\section*{Acknowledgments}
The authors would like to thank our many colleagues for discussions that seeded and developed the ideas behind this article. 
Tasnif Rahman for discussing an alternate concentration conversion type and community FBA.
T.J. Sego for many discussions on the role of protocols in multiscale modeling.
E.A. is funded by NSF award OCE-2019589 to the Center for Chemical Currencies of a Microbial Planet, by NIH award P41GM109824 to the Center for Reproducible Biomedical Modeling, and John Templeton Foundation award \#62825.

\section*{Author contributions}
E.A. and R.K.S. together conceived the framework, R.K.S. developed the process-bigraph and bigraph-schema software, E.A. developed the supplement's formal framework, the spatio-flux examples, viva-superpowers, wrote the manuscript, and made the figures.

\section*{Competing interests}
The authors declare no competing interests.

\section*{Data availability}
No datasets were generated or analyzed during this study.

\section*{Code availability}
All software is open-source and available at \url{https://github.com/vivarium-collective}.

\section*{Supporting information}

\paragraph*{S1 Text.}
\label{S1_Text}
{\bf Formal framework specification (Supplement~1).}
Technical underpinnings of the Process Bigraph framework.
(i) overview of the Vivarium~2.0 software suite;
(ii) schema-based semantics -- paths, schema trees and state trees, the type engine and type registry;
(iii) textual representations of a process bigraph;
(iv) links, ports, and the derived wiring map;
(v) processes and \emph{delta semantics} -- typed function signatures, type-directed update operators;
(vi) the tuple definition $B=(\Sigma,x,R_T,R_L)$ of a process bigraph;
(vii) composites, their \texttt{interface} and \texttt{bridge}, and hierarchical composition;
(viii) orchestration -- discrete event scheduling, update reconciliation, workflows, and the large-scale orchestration;
(ix) emitters and result capture;
(x) execution protocols.

\paragraph*{S2 Text.}
\label{S2_Text}
{\bf Spatio-Flux worked example (Supplement~2).}
A concrete Spatio-Flux instantiation of the Process Bigraph framework, mirroring the abstract semantics of S1~Text with executable process bigraphs.
(i) Spatio-Flux schema-level types (fields, particles, \texttt{conc\_counts}, complex particle records);
(ii) typed process and step definitions for field dynamics (diffusion/advection), stochastic and deterministic particle movement, and particle--field metabolic coupling;
(iii) composite documents and wiring diagrams for the multiscale examples used in the main text;
(iv) a matched comparison between Spatio-Flux and the COMETS Java backend (via \texttt{cometspy}~\cite{dukowski2021comets});
(v) runtime decomposition and wall-clock scaling results for local and distributed execution modes.

\nolinenumbers

\bibliography{references}

\end{document}